\documentclass[prx,aps,twocolumn,superscriptaddress]{revtex4-2}

\usepackage[pdftex]{graphicx}
\usepackage{amsbsy,amssymb,amsmath,bm,mathtools}
\usepackage{amsfonts}

\usepackage{xspace}
\usepackage{bm}
\usepackage{color}
 
\usepackage{url}
\usepackage[section]{placeins}
\usepackage[colorlinks=true,linkcolor=blue,citecolor=blue]{hyperref}

\usepackage[mathscr]{euscript}

\usepackage{graphicx}
\usepackage{dcolumn}

\begin{document}

\preprint{APS/123-QED}

\title{Machine-learning force-field models for dynamical simulations of metallic magnets}

\author{Gia-Wei Chern}
\email{gchern@virginia.edu}
\thanks{Corresponding author}
\affiliation{Department of Physics, University of Virginia, Charlottesville, VA 22904, USA}

\author{Yunhao Fan}
\affiliation{Department of Physics, University of Virginia, Charlottesville, VA 22904, USA}

\author{Sheng Zhang}
\affiliation{Department of Physics, University of Virginia, Charlottesville, VA 22904, USA}

\author{Puhan Zhang}
\affiliation{Department of Physics, University of Virginia, Charlottesville, VA 22904, USA}

\begin{abstract}
We review recent advances in machine-learning (ML) force-field methods for Landau-Lifshitz-Gilbert (LLG) simulations of itinerant electron magnets, focusing on their scalability and transferability. Built on the principle of locality, a deep neural-network model is developed to efficiently and accurately predict electron-mediated forces governing spin dynamics. Symmetry-aware descriptors constructed through a group-theoretical approach ensure rigorous incorporation of both lattice and spin-rotation symmetries. The framework is demonstrated using the prototypical s-d exchange model widely employed in spintronics. ML-enabled large-scale simulations reveal novel nonequilibrium phenomena, including anomalous coarsening of tetrahedral spin order on the triangular lattice and the freezing of phase-separation dynamics in lightly hole-doped, strong-coupling square-lattice systems. These results establish ML force-field frameworks as scalable, accurate, and versatile tools for modeling nonequilibrium spin dynamics in itinerant magnets.
\end{abstract}

\maketitle

\section{Introduction}

\label{sec:intro}

Itinerant frustrated magnets host a rich variety of spin and electronic textures stabilized by electron-mediated interactions. Among them, magnetic vortices and skyrmions have attracted particular attention for their topological stability and potential spintronic applications~\cite{bogdanov89,rossler06,muhlbaure09,yu10,yu11,seki12,nagaosa13}. The Berry phases associated with these textures act as emergent magnetic fields, giving rise to novel transport phenomena such as the anomalous Hall effect. Another striking example is the mixed-phase state in colossal magnetoresistance (CMR) manganites, where ferromagnetic metallic and antiferromagnetic or charge-ordered insulating regions coexist~\cite{dagotto_book,dagotto05,moreo99,dagotto01,mathur03,nagaev02,fath99,renner02,salamon01}. These nanoscale textures are highly responsive to external stimuli—magnetic fields, pressure, or doping—leading to dramatic transport changes.

Modeling the real-time dynamics of such complex spin textures remains computationally challenging. While classical spins evolve according to the Landau-Lifshitz-Gilbert (LLG) equation, the effective fields driving the dynamics originate from quantum electron–spin exchange interactions. Direct simulations are therefore prohibitively expensive, as each LLG step requires solving a many-body electronic problem. Machine learning (ML) offers a powerful alternative. As emphasized by Kohn~\cite{kohn96,prodan05}, linear-scaling electronic algorithms hinge on the principle of locality—a concept naturally realized in Behler-Parrinello (BP)-type ML force-field frameworks developed for {\em ab initio} molecular dynamics~\cite{behler07,bartok10,li15,shapeev16,behler16,botu17,smith17,zhang18,deringer19,mcgibbon17,suwa19,chmiela17,chmiela18,sauceda20,unke21} and extended to adiabatic lattice dynamics in condensed-matter systems~\cite{zhang22,zhang22b,cheng23,Ghosh24}.

Here, we present a scalable ML force-field framework for the adiabatic dynamics of itinerant magnets, generalizing the BP architecture~\cite{zhang20,zhang21,zhang23,Shi23,Fan24}. A central feature is the use of symmetry-preserving magnetic descriptors that remain invariant under lattice point-group and spin-rotation operations while being differentiable with respect to spin orientations. We demonstrate the method for the one-band s-d exchange model, widely used in spintronics. ML-trained force fields reproduce noncoplanar tetrahedral spin order and its anomalous coarsening dynamics on the triangular lattice~\cite{Fan24}, and reveal a correlation-induced freezing phenomenon in the phase-separated state of a lightly hole-doped, strong-coupling square-lattice system~\cite{zhang20}.

\section{Behler-Parrinello Force-field architecture}

\begin{figure*}[t]
\centering
\includegraphics[width=1.99\columnwidth]{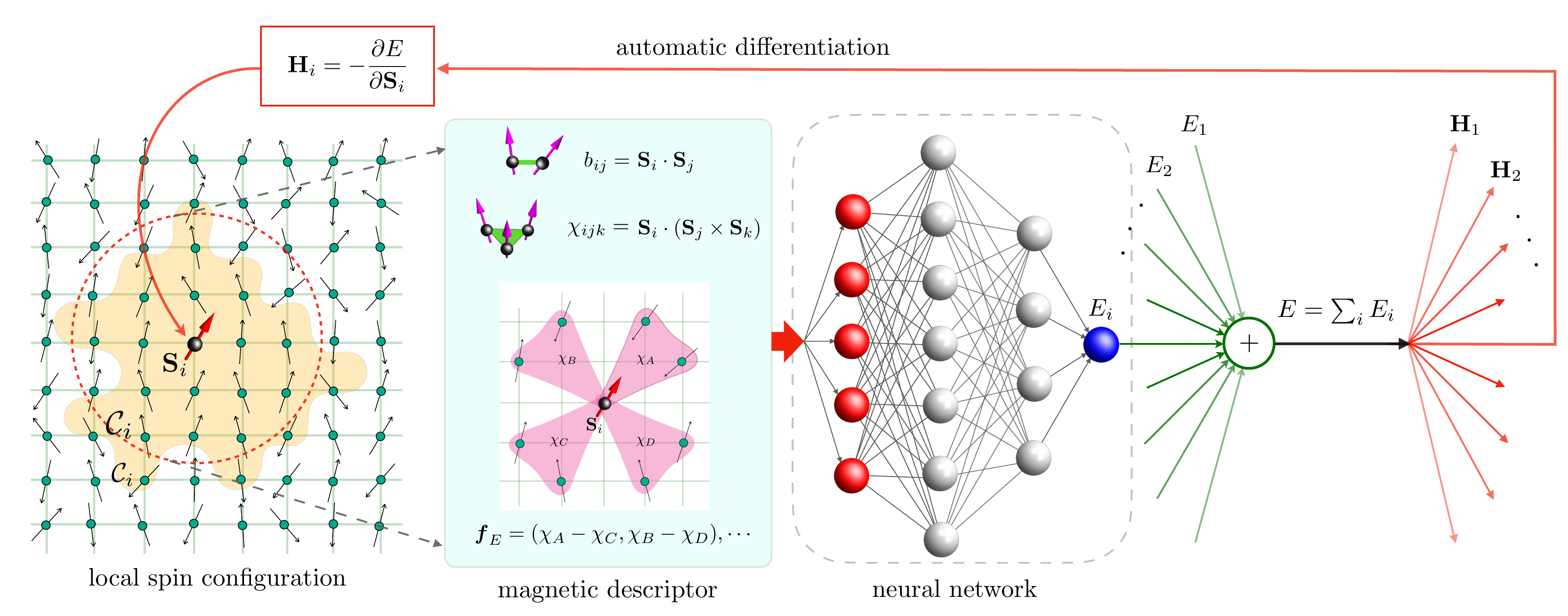}
\caption{Schematic diagram of ML force-field model for itinerant electron magnets. A descriptor transforms the neighborhood spin configuration $\mathcal{C}_i$ to effective coordinates $\{ p^\Gamma_r, \eta^\Gamma_r \}$ which are then fed into a neural network (NN). The output node of the NN corresponds to the local energy $\epsilon_i = \varepsilon(\mathcal{C}_i)$ associated with spin $\mathbf S_i$. The corresponding total potential energy $ E$ is obtained from the summation of these local energies. Automatic differentiation is employed to compute the derivatives $\partial E / \partial \mathbf S_i$ from which the local exchange fields $\mathbf H_i$ are obtained. }
    \label{fig:ml-scheme}
\end{figure*}

We consider a one-band s-d model, as a representative system for itinerant magnets, described by the following Hamiltonian
\begin{eqnarray}
	\label{eq:H1}
	\hat{\mathcal{H}} =- \sum\limits_{ ij , \alpha} t_{ij}  \left(\hat{c}_{i \alpha}^{\dagger}\hat{c}^{\,}_{j \alpha} + \mbox{h.c.} \right)
	-J \sum_{i, \alpha\beta} \mathbf {S}_i \cdot \hat{c}^\dagger_{i\alpha} \bm\sigma^{\,}_{\alpha\beta} \hat{c}^{\,}_{i \beta}.  \quad
\end{eqnarray}
Here $\hat{c}_{i \alpha}^{\dagger}(\hat{c}^{\,}_{i \alpha})$ is the creation (annihilation) operators of an electron with spin $\alpha = \uparrow, \downarrow$ at site-$i$. The first  term describes the Hamiltonian for the $s$-electrons, while $J$ in the second term represents the exchange coupling between local magnetic moment $\mathbf S_i$ from the $d$-electrons and spin of $s$-electrons. Here the local moments are approximated by classical spins with a fixed unit length.

The dynamics of the classical local spins is governed by the  Landau-Lifshitz-Gilbert equation
\begin{eqnarray}
	\label{eq:LLG}
	\frac{d\mathbf S_i}{dt} = \gamma \mathbf S_i \times ( \mathbf H_i + \bm\eta_i )  
	- \gamma \lambda \mathbf S_i \times \left[ \mathbf S_i \times ( \mathbf H_i + \bm\eta_i ) \right].
\end{eqnarray}
Here, $\gamma$ denotes the gyromagnetic ratio, $\lambda$ is a dimensionless damping parameter, $\mathbf H_i$ is the local effective field arising from the s-d exchange interaction with itinerant electrons, and $\bm\eta_i$ represents the stochastic magnetic field associated with thermal fluctuations modeled as a Gaussian random process.

Within the adiabatic approximation--analogous to the Born-Oppenheimer scheme in quantum MD~\cite{Marx09}, the local exchange force is given by the derivative of an effective spin-dependent energy $E = E(\{\mathbf S_i \})$:
\begin{eqnarray}
	\label{eq:H-force}
	\mathbf H_i = - \frac{\partial E}{\partial \mathbf S_i}
\end{eqnarray}
Formally, this effective energy can be obtained by integrating out electrons given a frozen spin configuration. For example, in the weak-coupling limit ($J \ll t_{ij}$), second-order perturbation theory yields an RKKY-type long-range spin-spin interaction~\cite{Ruderman1954,Kasuya1956,Yosida1957}. At intermediate to strong coupling, a systematic expansion is, however, exceedingly cumbersome and often impractical. The effective energy can be computed by numerically integrating out electrons on-the-fly,
\begin{eqnarray}
	\label{eq:E_quantum}
	E = \bigl\langle \hat{\mathcal{H}}(\{\mathbf S_i\}) \bigr\rangle 
	= {\rm Tr}\bigl(\hat{\rho}_e \hat{\mathcal{H}}  \bigr),
\end{eqnarray}
with $\hat{\rho}_e = e^{-\beta \hat{\mathcal{H}}}/\mathcal{Z}$ the equilibrium density matrix for a fixed spin configuration and $\mathcal{Z} = {\rm Tr} e^{-\beta \hat{\mathcal{H}}}$ the partition function. Since the Hamiltonian is quadratic in fermionic operators, exact diagonalization (ED) can be employed, but its $\mathcal{O}(N^3)$ scaling makes large-scale LLG simulations prohibitive. More efficient schemes, such as the linear-scaling kernel polynomial method (KPM), are available, though their implementation is intricate and they cannot treat systems with electron-electron interactions.

Machine learning (ML) methods provide a general linear-scaling approach to computing effective local fields~$\mathbf H_i$. As pointed out by W. Kohn, such scaling is enabled by the locality principle, or the ``nearsightedness'' of many-electron systems~\cite{kohn96,prodan05}. This principle implies that extensive quantities can be evaluated through a divide-and-conquer strategy.
Indeed, the locality principle underlies modern ML-based force-field methods that enable large-scale MD simulations with quantum accuracy~\cite{behler07,bartok10,li15,shapeev16,behler16,botu17,smith17,chmiela17,zhang18,chmiela18,deringer19,mcgibbon17,suwa19,sauceda20}.  A linear-scaling framework exploiting this principle was pioneered by Behler-Parrinello (BP)~\cite{behler07} and Bart\'ok {\em et al.}~\cite{bartok10}. Here, we extend the BP architecture to predict local exchange fields in itinerant spin systems.

A schematic of the ML force-field model is shown in Fig.~\ref{fig:ml-scheme}. The effective energy $E$, defined in Eq.~(\ref{eq:E_quantum}), is decomposed into local contribution $\epsilon_i$ associated with lattice site-$i$,
\begin{eqnarray}
	\label{eq:E_ML}
	E = \sum_i \epsilon_i = \sum_i \varepsilon(\mathcal{C}_i).
\end{eqnarray}
In the second expression, we invoke the locality principle, assuming that the site energy $\epsilon_i$ depends only on the local magnetic environment,  denoted as $\mathcal{C}_i$,  through a universal function $\varepsilon(\cdot)$ which is determined by the underlying electron Hamiltonian. In practical implementations, the neighborhood is defined as the collection of spins within a cutoff radius~$r_c$ around the $i$-th site: $\mathcal{C}_i = \{ \mathbf S_j \,  \bigl| \, |\mathbf r_j - \mathbf r_i| < r_c \}$. The complex functional dependence of the local energy on the magnetic environment is to be approximated by a deep-learning neural network. The effective local field $\mathbf H_i$, defined as the derivative of the total energy [Eq.~(\ref{eq:H-force})], can be efficiently computed using automatic differentiation~\cite{Paszke17,Baydin18}.

A key component of the ML force-field model is the representation, or descriptor, of local spin configurations~$\mathcal{C}_i$. In BP-type schemes, the output energy $\epsilon_i$ is a scalar and must remain invariant under the system’s symmetry operations. Thus, descriptors must both distinguish distinct spin configurations and preserve the underlying symmetries of the electron model. Descriptors play a similarly central role in BP-type force fields for quantum MD, where they are required to respect symmetry operations, such as translation, rotation, and reflection, of the 3D Euclidean group $E(3)$, as well as permutation of identical atoms~\cite{behler07,bartok10,li15,behler11,ghiringhelli15,bartok13,drautz19,himanen20,huo22}.

Here we employ group-theoretical methods to construct magnetic descriptors for the s-d model. In contrast to atomic descriptors in MD systems, where the relevant symmetry is the full $E(3)$ group, the lattice-based s-d model reduces this to the discrete translation group and the on-site point group~\cite{Ma19,Liu22,zhang22}. In addition, the model possesses a continuous symmetry: the s-d Hamiltonian is invariant under global SO(3)/SU(2) spin rotations~\cite{zhang21}. The local energy function $\varepsilon(\mathcal{C}_i)$ must respect both spin-rotation and lattice symmetries. To this end, we first note that spin-rotation invariance is ensured by expressing $\varepsilon(\mathcal{C}_i)$ in terms of bond variables $b_{jk}$ and scalar chiralities $\chi_{jmn}$,
\begin{eqnarray}
	\label{eq:bond-chirality}
	b_{jk} = \mathbf S_j \cdot \mathbf S_k, \qquad \chi_{jmn} = \mathbf S_j \cdot \mathbf S_m \times \mathbf S_n,
\end{eqnarray}
which correspond to two- and three-spin correlations.

For lattice symmetries, we first decompose the set of bond/chirality variables $\{ b_{jk}, \chi_{jmn} \}$ around site-$i$, which forms a basis of a high-dimensional representation of the on-site point group into irreducible representations (IRs). For instance, consider the four chirality variables ${\chi_A, \chi_B, \chi_C, \chi_D}$ in Fig.~\ref{fig:ml-scheme} for a square lattice with on-site symmetry group $D_4$. The decomposition is $4 = A_1 + B_1 + E$, with corresponding IR coefficients: $f^{A_1} = \chi_A + \chi_B + \chi_C + \chi_D$, $f^{B_1} = \chi_A - \chi_B + \chi_C - \chi_D$, and $\bm f^E = (\chi_A - \chi_C, \chi_B - \chi_D)$.  For each fundamental IR, we collect its component coefficients into a vector $\bm f^{(\Gamma,r)} = (f^{(\Gamma,r)}_1, \cdots, f^{(\Gamma,r)}_{n_\Gamma})$, where $\Gamma$ denotes the IR, $n_\Gamma$ its dimension, and $r$ the multiplicity. A basic class of invariants is the power spectrum,
\begin{eqnarray}
\label{eq:power-spectrum}
p^\Gamma_r = |\bm f^\Gamma_r|^2.
\end{eqnarray}
While the power spectrum ignores relative phases between IRs, these can be incorporated through bispectrum coefficients $b^{\Gamma, \Gamma_1, \Gamma_2}_{r, r_1, r_2}$~\cite{kondor07,bartok13}, constructed from triple products of IR coefficients and Clebsch-Gordon coefficients. Because IRs often have large multiplicities, bispectrum descriptors can be numerous and redundant.
To reduce complexity, we introduce reference IR coefficients $\bm f^\Gamma_{\rm ref}$~\cite{Ma19,zhang22,zhang21}, obtained from the IR decomposition of bond and chirality variables coarse-grained over large blocks. These define a phase variable,
\begin{eqnarray}
\label{eq:IR-phase}
\eta^\Gamma_r \equiv \bm f^\Gamma_r \cdot \bm f^\Gamma_{\rm ref} / |\bm f^\Gamma_r| |\bm f^\Gamma_{\rm ref}|,
\end{eqnarray}
allowing relative phases to be inferred consistently. The final feature set thus consists of amplitudes $p^\Gamma_r$, phases $\eta^\Gamma_r$, and a selected set of bispectrum coefficients obtained from the reference IR.  The site energy $\epsilon_i$, produced by the NN, depends only on these invariants, thereby preserving the full symmetry of the s-d model.

\begin{figure}[t]
\centering
\includegraphics[width=0.92\columnwidth]{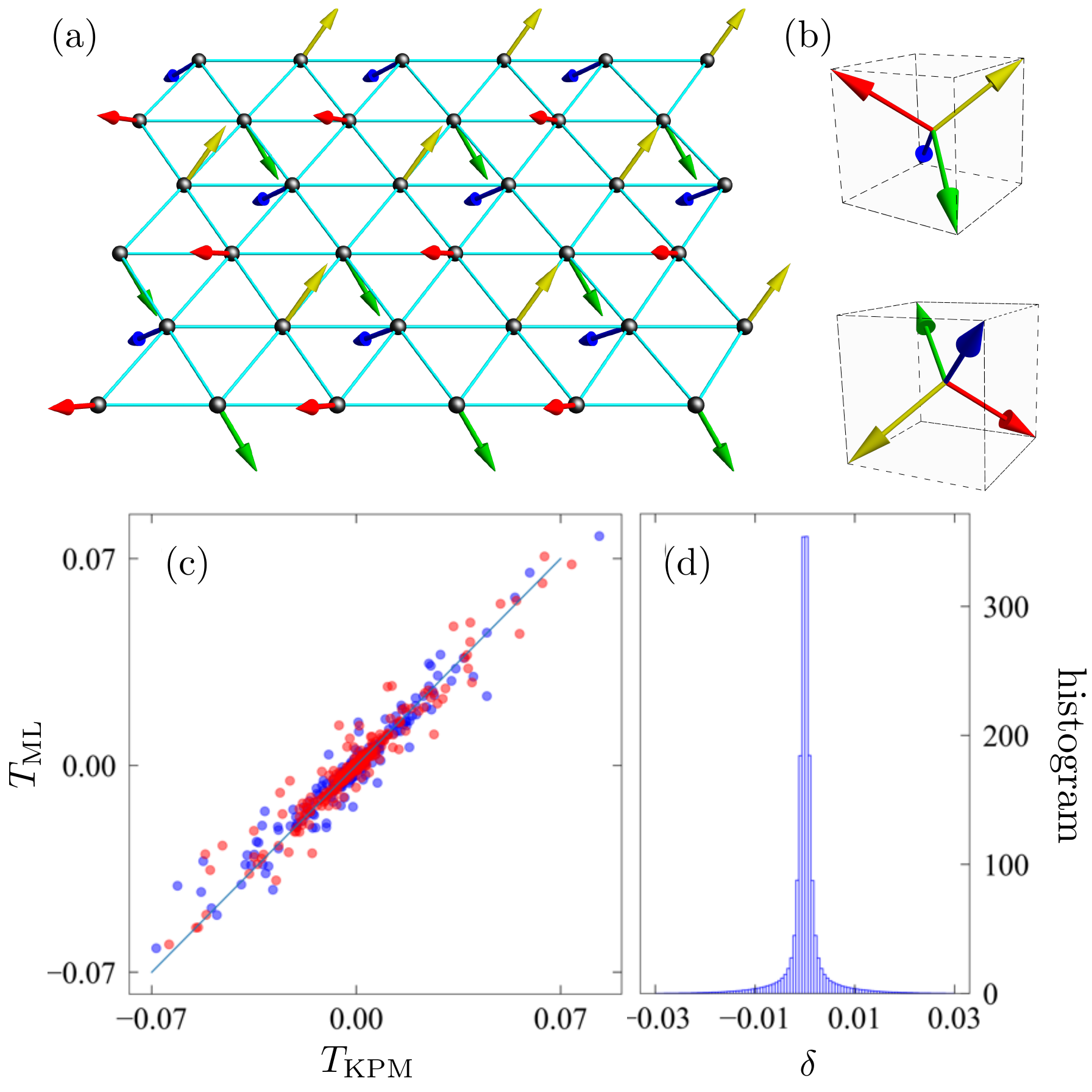}
\caption{(a) Tetrahedral spin order on a triangular lattice [Eq.~(\ref{eq:triple-Q})]. (b) Two inequivalent spins in a quadrupled unit cell with opposite chirality. (c) Comparison of ML-predicted torque $T_{\rm ML}$ with KPM results $T_{\rm KPM}$, where $\mathbf T_i = \mathbf S_i \times \mathbf H_i$. (d) Histogram of prediction error $\delta = T_{\rm KPM} - T_{\rm ML}$.}
    \label{fig:chiral-order}
\end{figure}

As discussed above, one of the major advantages of the proposed scalable ML framework is its dramatically enhanced computational efficiency--both in terms of algorithmic complexity and actual runtime. For instance, a 10,000-time-step simulation of a $50\times50$ system using the exact-diagonalization (ED) method requires approximately 20 CPU hours. While limited parallelization on GPU can offer a modest improvement, the ED calculation based on the LAPACK package is inherently non-parallelizable. In contrast, the same simulation performed with the ML force-field model takes only about 5 minutes, representing nearly a three-order-of-magnitude ($\sim$1000-fold) speedup. This dramatic improvement stems from the difference in computational scaling---$\mathcal{O}(N^3)$ for the ED approach versus linear $\mathcal{O}(N)$ scaling for the ML model. 
Even when compared with the KPM, which also exhibits linear scaling and can be efficiently implemented on GPU, the ML framework remains superior. A 500-step simulation of a $96\times96$ system takes about 100 minutes using KPM, whereas the same run with the ML model requires only around 20 minutes, demonstrating an additional $\sim$5-fold gain in efficiency.

\section{Chiral domain coarsening of tetrahedral spin order }

\begin{figure*}[t]
\centering
\includegraphics[width=1.99\columnwidth]{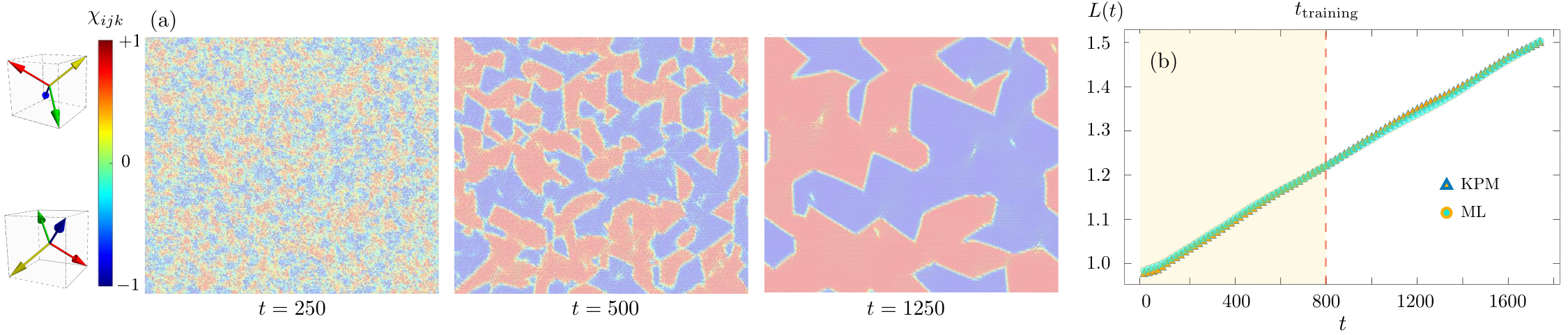}
\caption{(a) Snapshots of local scalar chirality $\chi_{ijk} = \mathbf S_i \cdot \mathbf S_j \times \mathbf S_k$, with $(ijk)$ denoting sites of an elementary triangle, at different times after a thermal quench of the triangular s–d model near $n=1/4$. (b) Characteristic length $L$ versus time, extracted from the chirality structure factor (linear scales on both axes). Yellow triangles and green circles show data from LLG quenches using KPM and ML-calculated fields, respectively. The ML model was trained only on KPM–LLG data up to $t_{\rm training} = 800$, marked by the yellow shaded region.}
    \label{fig:chiral-coarsening}
\end{figure*}

We first apply the above ML framework to the nearest-neighbor triangular-lattice s-d model at electron filling fraction $f \sim 1/4$. Notably, the ground state at this filling fraction as well as $f = 3/4$  is a non-coplanar tetrahedral order with a quadrupled unit cell~\cite{Martin2008,Akagi10,Kato2010,Azhar17,Chern2012}. In this order, the local spins point toward the corners of a regular tetrahedron; see Fig.~\ref{fig:chiral-order}(a) and (b). More generally, magnetic orders with a quadrupled unit cell on the triangular lattice can be described by a triple-$\mathbf Q$ structure,
\begin{eqnarray}
	\label{eq:triple-Q}
	\mathbf S_i = \bm \Delta_1 e^{i \mathbf M_1 \cdot \mathbf r_i} + \bm \Delta_2 e^{i \mathbf M_2 \cdot \mathbf r_i} +
	\bm \Delta_3 e^{i \mathbf M_3 \cdot \mathbf r_i},
\end{eqnarray}
with ordering vectors $\mathbf M_1 = (2\pi/a, 0)$ and $\mathbf M_{2,3} = (-\pi/a, \pm \sqrt{3}\pi/a)$ at the midpoints of the hexagonal Brillouin zone (BZ) edge, and vector order parameters $\bm \Delta_\eta$. The symmetric tetrahedral state corresponds to $\bm \Delta_\eta = \Delta \hat{\mathbf e}_\eta$, where $\Delta$ is the amplitude and $\hat{\mathbf e}_\eta$ ($\eta = 1, 2, 3$) are three orthogonal unit vectors, yielding spins at the corners of a regular tetrahedron.

The tetrahedral order is further characterized by a discrete $Z_2$ symmetry associated with the handedness of the spin texture. The scalar chirality on a triangular plaquette is defined as $\chi_{ijk} = \mathbf S_i \cdot \mathbf S_j \times \mathbf S_k$, which takes the uniform value $\chi_{ijk} = \pm 4 S^3 / 3\sqrt{3}$ in the perfect tetrahedral state. The two signs correspond to an Ising-like $Z_2$ symmetry spontaneously broken at low temperatures. 
The associated electronic state exhibits striking transport and magnetoelectric properties. In particular, the chiral spin order produces a quantized Hall conductivity $\sigma_{xy} = \pm e^2/h$ without external magnetic fields~\cite{Martin2008,Chern2010}.

Although long-range tetrahedral order is destroyed by thermal fluctuations in 2D, the $Z_2$ chirality order survives at finite temperature. Monte Carlo simulations of the triangular s-d model at $n \approx 1/4$~\cite{Kato2010} revealed a chiral phase transition that, despite its Ising-like order parameter, is strongly first order. This deviation from the expected 2D Ising universality suggests that the coarsening dynamics of chiral domains may also differ from the Allen-Cahn growth law expected for 2D Ising model. To investigate this, we built a ML force field model for the triangular-lattice s-d model with Hund's rule coupling $J = 3$, and a chemical potential $\mu = -3.2$, corresponding to a filling fraction $n \approx 1/4$; here both energies are measured in units of the nearest-neighbor hopping $t_{\rm nn}$. The training dataset were obtained from KPM method~\cite{Weisse06,Barros13,Wang18}. 

The neural network was implemented in PyTorch~\cite{Paszke2019} and consists of eight hidden layers with $2048\times1024\times512\times256\times128\times64\times64\times64$ neurons. The input layer contains 1806 nodes, corresponding to the feature variables ${G^\Gamma_r}$, while the single output neuron predicts the local energy. Rectified Linear Unit (ReLU) activations~\cite{Nair2010} are used between layers, and the network is optimized using the Adam algorithm~\cite{Kingma2014}. Since the torque $\mathbf{T}_i = \mathbf{S}_i \times \mathbf{H}_i$ is the primary driving force in LLG spin dynamics, the loss function incorporates the mean-square errors (MSE) of both the local torques and total energy: $ L = \sum_{i = 1}^N \left| \mathbf T_i - \hat{\mathbf T}_i \right|^2 + \eta_E \left| E - \sum_{i=1}^N \hat{\epsilon}_i \right|^2$, where quantities with hats denote ML predictions and $\eta_E$ controls the relative weight of the energy term. Two models were trained: one using only the torque loss ($\eta_E = 0$) and another with $\eta_E = 0.01$. The benchmark of the $\eta_E = 0$ model is shown in Fig.~\ref{fig:chiral-order}(c,d), the predicted forces agree well with the exact results, achieving an MSE of $1.64\times10^{-5}$ per spin without signs of overfitting. Additional dynamical benchmarks are provided in Ref.~\cite{Fan24}. We also note that although both achieve similar MSE values, including the energy term generally makes training more difficult. Since LLG simulations depend directly on the predicted torques, it is practical to begin training with the torque-only loss and gradually introduce the energy term by increasing $\eta_E$ in later epochs. The training of the the above NN models with 8 hidden layers took about $\sim 5$ days to reach the reported MSE.

Next we applied the trained ML model to perform large-scale thermal quench simulations of the s-d model. An initial state with random spins was suddenly coupled to a thermal bath at low temperature $T = 0.01$. Fig.~\ref{fig:chiral-coarsening}(a) shows snapshots of the local scalar chirality $\chi_{\triangle}(\mathbf r)$ at successive times following a thermal quench at $t=0$, with color indicating the chirality magnitude. Initially, the random spin configuration produces a disordered chirality pattern. As relaxation proceeds, domains of uniform chirality emerge and grow. These domains are separated by sharp interfaces, only a few lattice constants wide, across which the chirality vanishes.

The characteristic size $L(t)$ of chiral domains is obtained from the inverse width $\Delta q^{-1}$ of the $\mathbf q=0$ peak in the ensemble-averaged chirality structure factor. Fig.~\ref{fig:chiral-coarsening}(b) compares $L(t)$ from LLG simulations using ML- and KPM-calculated fields, showing excellent agreement and confirming that the ML force field accurately reproduces the phase-ordering dynamics. Notably, although the ML model was trained only on KPM-LLG data up to $t_{\rm training}=800$, it continues to predict $L(t)$ reliably at much longer times, demonstrating strong extrapolation beyond the training window.

More importantly, we find that the chiral domain size grows linearly with time, $L(t) \sim L_0 + a(t - t_0)$, in sharp contrast to the Allen–Cahn law $L \sim \sqrt{t}$ expected for a non-conserved Ising order parameter. Since the Allen–Cahn behavior arises from curvature-driven domain motion, the observed linear growth likely reflects the distinctive domain morphology. As shown in Fig.~\ref{fig:chiral-coarsening}(a), late-stage chiral domains exhibit nearly straight interfaces aligned with the principal symmetry directions of the triangular lattice. The zero curvature of these straight boundaries implies vanishing domain-wall motion under the Allen–Cahn mechanism. In this case, domain growth is most likely governed by the dynamics of sharp corners, suggesting the need for a theoretical framework of chiral-domain coarsening based on corner motion as point defects.

\section{Phase separation dynamics of double-exchange models}

In this section, we apply the ML force-field framework to study phase-separation dynamics in the strong-coupling, or double-exchange~\cite{zener51,anderson55,degennes60}, regime of the s-d model. Phase separation phenomena play a central role in the behavior and functionality of many strongly correlated electron systems~\cite{schulz89,emery90,white00,tranquada95,dagotto_book,dagotto05,moreo99,dagotto01,mathur03,nagaev02}.
A prominent example is the complex inhomogeneous states observed in manganites and magnetic semiconductors exhibiting the colossal magnetoresistance (CMR) effect~\cite{dagotto_book,dagotto05,moreo99,dagotto01,mathur03,nagaev02}. These nanoscale textures arise from the segregation of hole-rich ferromagnetic clusters within half-filled antiferromagnetic domains~\cite{fath99,renner02,salamon01}. Because the number of doped carriers is conserved, the segregation dynamics follow the classic theory of Lifshitz, Slyozov, and Wagner (LSW)~\cite{lifshitz61,wagner61}, which predicts a coarsening exponent $\alpha = 1/3$.

The square-lattice double-exchange model has been extensively studied theoretically~\cite{varma96,yunoki98,dagotto98,chattopadhyay01,pekker05}. In the regime of slight hole doping from half-filling, a mixed-phase state -- consisting of hole-rich ferromagnetic puddles embedded in a half-filled antiferromagnetic insulator -- emerges as a stable thermodynamic phase at strong Hund’s coupling~\cite{yunoki98,dagotto98,chattopadhyay01}. Despite extensive research on the static properties of such mixed-phase states in CMR materials, the kinetics of electron-correlation-driven phase separation remain largely unexplored. Fundamental questions, such as whether the process exhibits dynamical scaling and whether late-stage domain growth follows the LSW power law, are still open. The superior scalability of the ML force-field framework provides a powerful means to address these questions through large-scale simulations of phase separation in this model.

\begin{figure}
\centering
\includegraphics[width=0.95\columnwidth]{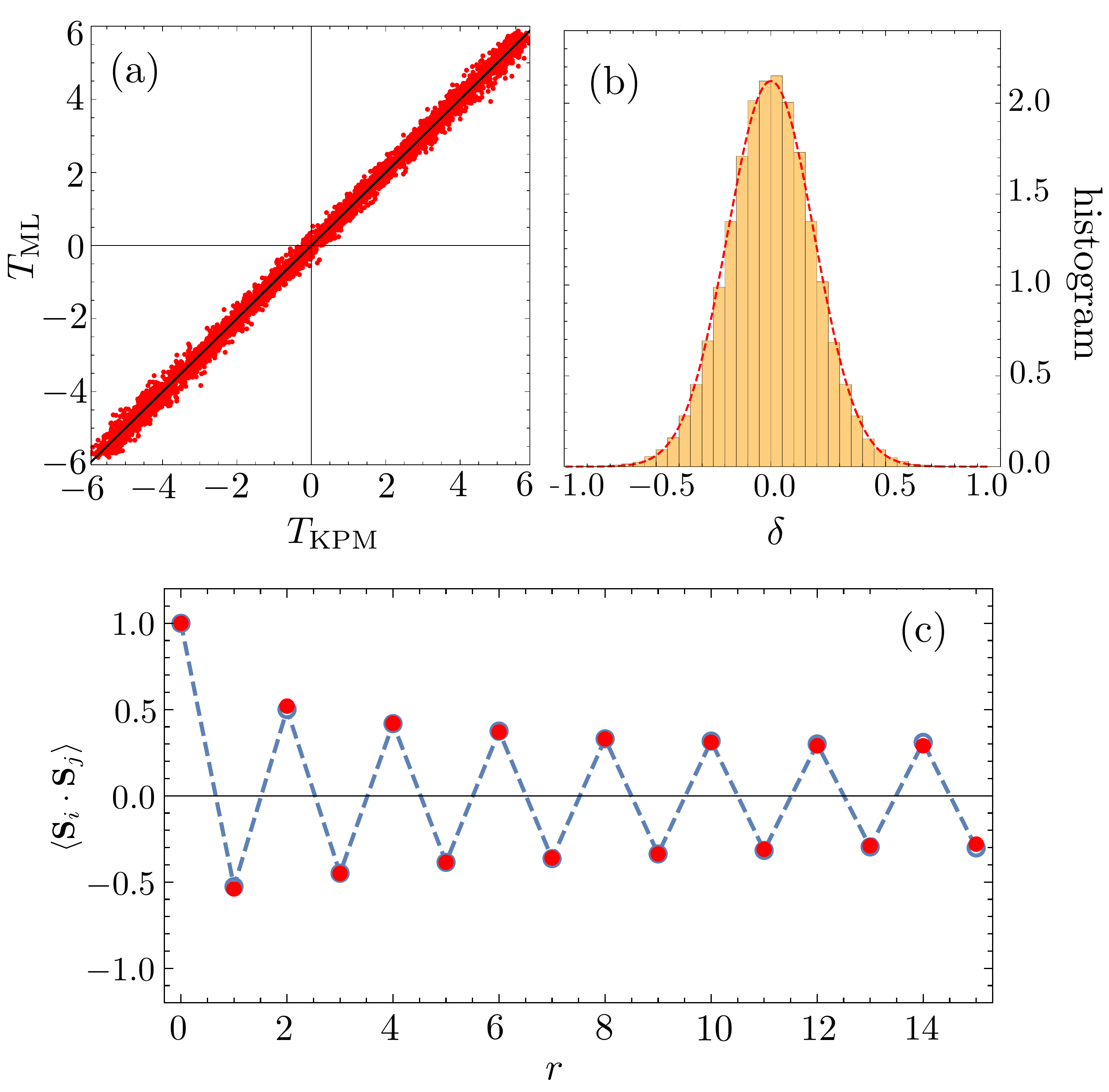}
\caption{(a) ML-predicted exchange fields versus exact results from the test dataset. (b) The distribution of the force difference $\delta = H_{\text{ML}} - H_{\text{exact}}$, well fitted by a normal distribution (red line) with variance $\sigma^2 = 0.035$. (c) Spin-spin correlation $\langle \mathbf S_i \cdot \mathbf S_j \rangle$ as a function of distance $r_{ij} = |\mathbf r_j - \mathbf r_i|$ along the $x$ direction at filling $f = 0.485$. Red dots show LLG results using NN models {without} Langevin noise, while blue lines correspond to ED-LLG simulations at $T = 0.022$.}
    \label{fig:ps-benchmark}
\end{figure}

\begin{figure*}[t]
\centering
\includegraphics[width=1.99\columnwidth]{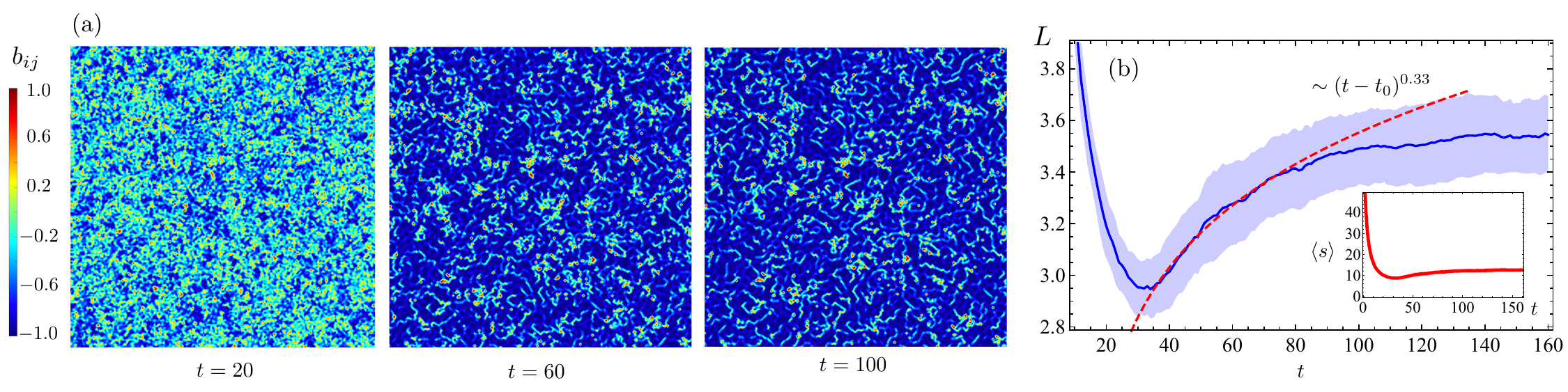}
\caption{(a) Density maps of local bond variables $b(\mathbf r_i)$ at four different times during the ML–LLG simulation on a $100\times100$ lattice with $1.5\%$ hole doping. (b) Average linear size $L = \langle s \rangle^{1/2}$ of FM clusters versus time after a thermal quench. The dash-dotted line denotes the $t^{1/3}$ power-law growth, while the dashed line shows a sublogarithmic dependence $L(t) \sim (\log t)^{\beta}$ with $\beta = 0.11$. The inset plots the time evolution of the average cluster size $\langle s \rangle$. }
    \label{fig:phase-separation}
\end{figure*}

We trained a six-layer neural network (NN) using PyTorch~\cite{Paszke2019,Nair2010} on 3500 snapshots of spin configurations and local exchange forces from ED–LLG simulations on a $30\times30$ lattice. Fig.~\ref{fig:ps-benchmark}(a) compares NN-predicted exchange fields $\mathbf H_i$ with exact results, while the deviation $\delta = H_{\text{ML}} - H_{\text{exact}}$ follows a Gaussian distribution with variance $\sigma^2 = 0.035$ [Fig.~\ref{fig:ps-benchmark}(b)]. The normal distribution of $\delta$ suggests that the ML uncertainty may act as an effective temperature in Langevin dynamics. We then combined the trained NN with LLG dynamics to simulate a thermal quench from random spins at $T = 0.022$ on a $30\times30$ lattice, using identical parameters as in the ED reference. The resulting spin–spin correlations, shown in Fig.~\ref{fig:ps-benchmark}(c) for filling $f = 0.485$ (1.5\% hole doping), agree closely with ED-LLG results, confirming the accuracy and dynamical transferability of the ML force field.

Large-scale quench simulations were performed using the trained ML model on a $100\times100$ lattice. Figure~\ref{fig:phase-separation}(a) shows snapshots of the local spin correlation, defined as the average over four nearest-neighbor bonds, $b_i = (\mathbf S_i \cdot \mathbf S_{i+\mathbf x} + \mathbf S_i \cdot \mathbf S_{i-\mathbf x} + \mathbf S_i \cdot \mathbf S_{i+\mathbf y} + \mathbf S_i \cdot \mathbf S_{i-\mathbf y})/4$. Positive (negative) $b_i$ corresponds to ferromagnetic (FM) [antiferromagnetic (AFM)] regions. The ML–LLG simulations reveal relaxation into an inhomogeneous state comprising extended AFM domains interspersed with small FM clusters.

We next analyze the kinetics of FM-domain growth. A FM cluster is defined as a connected region where all nearest-neighbor bonds exceed $b_{\rm th} = 0.5$. The time evolution of the characteristic length $L$ is shown in Fig.~\ref{fig:phase-separation}(b). Because the number of doped holes is conserved, the coarsening of such conserved fields follows the Cahn–Hilliard equation~\cite{cahn58,hohenberg77} or LSW theory~\cite{lifshitz61,wagner61}, predicting $L(t) \sim t^{1/3}$. However, as shown in Fig.~\ref{fig:phase-separation}(b), this scaling holds only at early times; at later stages, the domain size $L$ grows much more slowly than predicted by LSW theory.

The LSW theory describes diffusive interactions among minority-phase domains, where larger clusters grow by absorbing material from smaller ones via an evaporation–condensation mechanism. In our case, this corresponds to the migration of doped holes from small to large FM clusters. The early-stage $t^{1/3}$ scaling likely reflects this process. As AFM correlations develop in the half-filled background, however, each doped hole becomes surrounded by parallel spins through the double-exchange mechanism, forming a self-trapped FM cloud. This localization suppresses hole evaporation and halts the coarsening process, leading to a breakdown of the LSW scenario.

\section{Conclusion and Outlook}

In this paper, we have presented a scalable ML force-field framework for LLG simulations of itinerant electron magnets. Based on the principle of locality, the method expresses effective spin-dependent forces as functions of local spin environments using group-theoretical descriptors that rigorously preserve both lattice and spin symmetries. Analogous to the BP approach for ab initio MD simulations, the neural-network model predicts local exchange fields as energy gradients, achieving linear-scaling efficiency and high transferability across spin configurations. Applications to the prototypical s–d exchange model demonstrate the versatility of this framework and reveal nonequilibrium dynamical phenomena beyond the reach of conventional methods.

While this study focuses on the BP-type formulation, other ML architectures provide complementary and often more flexible approaches. Convolutional neural networks (CNNs), with their intrinsic local connectivity, naturally implement the principle of locality and enable scalable ML force fields for spin systems~\cite{cheng23b,tyberg25}. Recent advances further introduce elegant ways to embed both symmetry and scalability: equivariant neural networks enforce symmetry constraints directly within the architecture~\cite{thomas2018,anderson2019,townshend2020}, while graph neural networks (GNNs) achieve linear scaling via localized message passing and symmetry-preserving transformations of equivalent graph elements~\cite{scarselli09,reiser2022,batzner2022,batatia2025}. Extending these advanced architectures to model spin dynamics in itinerant magnets represents a promising avenue for future work.

\bigskip

\noindent {\bf Data availability} The data that support the findings of this study are openly available in \texttt{https://github.com/cherngroupUVA/ML\_double\_exchange} where one can find C and Python codes, trained model samples and data samples to successfully run the machine learning spin dynamics and reproduce results discussed in the main text.

\bigskip

\begin{acknowledgments}
This work was supported by the US Department of Energy Basic Energy Sciences under Contract No. DE-SC0020330. The authors acknowledge Research Computing at The University of Virginia for providing computational resources and technical support that have contributed to the results reported within this publication. 
\end{acknowledgments}

\bibliography{ref.bib}

\end{document}